\documentclass[aps,prd,preprint,amssymb,cite,
amsfonts,
nofootinbib,superscriptaddress]{revtex4}

\usepackage{graphicx}
\usepackage{bm,latexsym,amsmath,amssymb,amsfonts}
\usepackage[usenames,dvipsnames]{color}
\usepackage[colorlinks=true,linkcolor=blue,citecolor=blue,urlcolor=blue]{hyperref}
\usepackage{color}
\usepackage{soul}
\soulregister\cite7
\soulregister\ref7
\soulregister\pageref7
\usepackage{epsfig}
\usepackage{braket}
\usepackage[mathscr]{eucal}
\usepackage{cancel}
\usepackage{mathrsfs}
\usepackage{pgf, tikz}
\usetikzlibrary{shapes.geometric, arrows, positioning, calc, backgrounds}
\usepackage{slashed}
\usetikzlibrary{arrows,automata}
\usepackage{natbib}
\usepackage{pgfplots}
\pgfplotsset{compat=newest}

\definecolor{mypink1}{rgb}{0.858, 0.188, 0.478}
\definecolor{mypink2}{RGB}{219, 48, 122}
\definecolor{mypink3}{cmyk}{0, 0.7808, 0.4429, 0.1412}
\definecolor{mygray}{gray}{0.6}
\definecolor{pptbg}{rgb}{0.961,0.945,0.863}

\newcommand{\be}[1]{\begin{equation} \label{#1}}
\newcommand{\ee}{\end{equation}}
\newcommand{\bea}{\begin{eqnarray}}
\newcommand{\bean}{\begin{eqnarray*}}
\newcommand{\eea}{\end{eqnarray}}
\newcommand{\eean}{\end{eqnarray*}}
\newcommand{\ba}{\begin{array}}
\newcommand{\ea}{\end{array}}
\newcommand{\bel}{\begin{align}}
\newcommand{\eel}{\end{align}}
\newcommand{\nn}{\nonumber}

\newcommand{\doi}[1]{\href{#1}{\color{blue}{#1}} }

\newcommand{\rd}{\mathrm{d}}

\newcommand{\cA}{\mathcal{A}}
\newcommand{\cB}{\mathcal{B}}

\begin{document}
\title{Emergence of Universal Schwarzian Structure in Kerr Geometry}
\author{Hyeong-Chan Kim}
\affiliation{School of Liberal Arts and Sciences, Korea National University of Transportation, Chungju 380-702, Korea}

\email{hckim@ut.ac.kr}

\begin{abstract}
We show that, within a broad stationary-axisymmetric class, Kerr-type separability and hidden symmetry arise as a local consequence of the Einstein equations.
Without assuming separability, algebraic speciality, Killing--Yano symmetry, or global boundary conditions, we analyze stationary and axisymmetric geometries in a locally non-rotating orthonormal frame and impose a minimal local equilibrium condition, namely the absence of mixed energy-momentum fluxes. 
We find that the mixed Einstein equations enforce a rigid projective alignment between radial and angular sectors, uniquely characterized by a constant-Schwarzian constraint.
This constraint yields a universal classification of local solutions into M\"obius, exponential, and trigonometric branches, of which global regularity selects precisely the Kerr-type sector. 
In this sense, the kinematical core of Kerr geometry is already fixed locally, and the Schwarzian structure provides the local origin of Kerr rigidity.
\end{abstract}
\maketitle

\section{Introduction} \label{sec:intro}
The Kerr metric~\cite{Kerr1963} is the canonical rotating solution of the Einstein equations, the universal exterior geometry of astrophysical black holes, and one of the most rigid structures in classical gravity. 
It is distinguished by hidden symmetries responsible for the Carter constant, complete integrability, and separability of geodesic and wave equations~\cite{Carter1968,WalkerPenrose1970,FrolovKubiznak2007,Plebanski76,DietzRuediger1981}, while its near-extremal throat exhibits a universal low-energy sector closely related to near-horizon Kerr/near-AdS$_2$ dynamics~\cite{Bardeen:1999px,Guica:2008mu,Castro:2010fd} and of direct relevance to the long-lived ringdown response of rapidly rotating black holes~\cite{Detweiler:1980gk,Yang:2013uba}.
These features are usually regarded either as special properties of the Kerr solution itself, tied to Petrov type~D and Killing-Yano structure~\cite{Goldberg62,Kubinznak2007,Frolov2007,Santillan:2011sh}, or as consequences of global assumptions entering black-hole uniqueness theorems~\cite{Carter71,Robinson1975,Mazur2000,Chrusciel2012}. 

A more local question, however, is whether the Einstein equations already encode the kinematical core of this rigidity before asymptotic flatness, horizon regularity, or separability are imposed. 
In this paper, we show that, within a broad class of stationary axisymmetric geometries, the mixed Einstein equations enforce a nontrivial \emph{projective-Schwarzian constraint} that rigidly aligns the radial and angular sectors. 
The appearance of the Schwarzian derivative is especially notable because the same structure governs the universal boundary reparametrization sector of nearly-AdS$_2$ gravity and related low-energy holographic systems~\cite{Maldacena2016,Kitaev2018}.
This identifies a local invariant precursor of Kerr-type separability and hidden symmetry, while also revealing the same geometric structure in the universal near-horizon sector relevant for near-AdS$_2$ physics and near-extremal gravitational ringdown.

To address this, we consider a stationary and axisymmetric metric ansatz~\eqref{metric:gen} and impose a \emph{local equilibrium condition}: the Einstein tensor is diagonal in a locally non-rotating orthonormal frame,
\begin{equation}
G_{\hat a\hat b}=0, \qquad a\neq b .
\label{key}
\end{equation}
Importantly, this condition does not assume vacuum; it remains compatible with any comoving stress tensor whose mixed components vanish in the locally non-rotating frame.
For the ansatz of interest, all mixed components vanish identically except $G_{\hat0\hat3}$ and $G_{\hat1\hat2}$, so the nontrivial content of Eq.~\eqref{key} reduces to this pair alone.

Our main result is that these two local conditions already force the geometry into a projectively rigid and separable form. 
The reduced Einstein system organizes into a characteristic Riccati equation whose closure leads to a compatibility condition equivalent to equality of Schwarzian derivatives between the radial and angular sectors,
\begin{equation}
\{\Gamma,R\}=\{p,z\}=C .
\end{equation}
This constant-Schwarzian condition yields three universal local branches, M\"obius, exponential, and trigonometric, distinguished by the sign of the invariant $C$. 
After imposing standard global regularity and periodicity conditions, the trigonometric branch is generically excluded, leaving the Kerr-compatible sector, within which Petrov type~D and Killing--Yano structure arise naturally.
In this sense, hidden symmetry appears not as an accidental property of a special exact solution, but as a consequence of a local projective rigidity encoded in the Einstein equations.
This projective-Schwarzian constraint is moreover robust under a broader redistribution of radial and angular warp factors, as shown in App.~\ref{App:robustness}, indicating that it is not an artifact of the specific parametrization adopted here.

The result is established within the stationary-axisymmetric class introduced in Sec.~\ref{sec:ansatz}, where the $(t,\phi)$-fiber functions are taken to depend on a single variable in each sector.

The appearance of the Schwarzian derivative also suggests a broader connection to the universal structures familiar from nearly-AdS$_2$ gravity and related low-energy systems, where Schwarzian dynamics governs boundary reparametrization modes.
In the Kerr context, this points to a natural near-horizon interpretation of the projective invariant, without implying that the full four-dimensional dynamics reduces to a two-dimensional theory.
It also suggests a corresponding role for the same invariant in the near-extremal wave dynamics of Kerr, providing a possible geometric link between local hidden symmetry and the universal low-frequency behavior of rapidly rotating black holes.

Our analysis therefore identifies a local precursor to Kerr rigidity that is both mathematically sharp and physically broad. 
Rather than obtaining Kerr structure only after imposing global boundary conditions, we show that the Einstein equations together with the local equilibrium condition already enforce its kinematical core within the present class: projective alignment, separability, Petrov type~D, and the associated hidden symmetry.

The remainder of this paper is organized as follows.
In Sec.~\ref{sec:ansatz}, we introduce the stationary and axisymmetric ansatz and the local equilibrium condition.
In Secs.~\ref{sec:char-var} and \ref{sec:Riccati}, we show how the mixed Einstein equations reduce to a characteristic system and then to a Riccati equation.
In Sec.~\ref{sec:schwarzian}, we derive the constant-Schwarzian constraint, establish the projective alignment between the radial and angular sectors, and classify the resulting local branches.
We conclude with a summary and discussion, while technical details are collected in the appendices.

\vspace{.3cm}
\section{Stationary and Axisymmetric Ansatz and Local Equilibrium} \label{sec:ansatz}
We consider a general stationary and axisymmetric spacetime and formulate the Einstein equations in an orthonormal (non-coordinate) frame adapted to these symmetries.
In such a frame, mixed components of the Einstein tensor have a direct physical interpretation as local energy and momentum fluxes, rather than coordinate artifacts.
This allows us to impose a physically meaningful local equilibrium condition in a locally non-rotating frame.
Guided by stationarity and axial symmetry, we adopt the metric ansatz taken in Ref.~\cite{Kim:2026gog}
\bea
ds^2 &=&
 -  \frac{  \Sigma \Delta}
	{ q \left(\Gamma -   a^2p \right)^2 }
	\left( dt - a p d\phi \right)^2 
+ \frac{1}{q} \frac{\Sigma}{\Delta }dr^2 
 + \Sigma d\theta^2 
 + \frac{  \Sigma \sin^2\theta }
	{  \left(\Gamma -  a^2p \right)^2 }
	(\Gamma d\phi -a dt)^2 ,
\label{metric:gen}
\eea
where $\Gamma=\Gamma(r)$, $p=p(x)$, $q=q(x)$, and $\Sigma=\Sigma(r,x)$ with $x=\cos\theta$, while $\Delta=\Delta(r)$ is left arbitrary.
Except that fiber sector functions are separately radial/angular, this ansatz is sufficiently general to encompass Kerr-type geometries without assuming separability, Petrov type, or hidden symmetries \emph{a priori}.
The ansatz simply parametrizes a broad class of stationary axisymmetric geometries with independent radial and angular functions; the Kerr-type structure emerges only after imposing the local equilibrium conditions~\eqref{key}.
We emphasize that Eq.~\eqref{metric:gen} captures a broad class of stationary axisymmetric metrics with a warped two-plane base and single-variable $(t,\phi)$-fiber functions; more general geometries with nontrivial $r$-$\theta$ mixing in the fiber sector (e.g., $\Gamma=\Gamma(r,\theta)$) or with genuinely non-removable $g_{r\theta}$ are outside our present scope.

A convenient orthonormal coframe $\{e^{\hat a}\}$ satisfying
$ds^2=\eta_{\hat a\hat b}e^{\hat a}e^{\hat b}$ with
$\eta_{\hat a\hat b}=\mathrm{diag}(-1,1,1,1)$ is
\bea
e^{\hat 0} &=& 
\sqrt{\frac{  \Sigma \Delta}{ q}} 
\frac{dt - a p \,d\phi }{ \Gamma -   a^2p  },\qquad
e^{\hat 1} =\sqrt{\frac{  \Sigma }{q\Delta}}  \,dr, \nonumber\\
e^{\hat 2} &=&\sqrt{\Sigma } \,d\theta,\qquad
e^{\hat 3} =
\frac{  \sqrt{\Sigma } \sin\theta}{\Gamma- a^2 p}\, (\Gamma \,d\phi - a\, dt).
\label{vierbein}
\eea
Any other coframe related by a local Lorentz transformation yields equivalent tensor components.

We impose the local equilibrium condition~\eqref{key} that the Einstein tensor be diagonal in this locally non-rotating orthonormal frame.
For the ansatz of interest, all mixed components vanish identically except $G_{\hat0\hat3}$ and $G_{\hat1\hat2}$, so the nontrivial content of Eq.~\eqref{key} reduces to this pair alone.
Importantly, this condition does not assume vacuum; it remains compatible with any comoving stress tensor whose mixed components vanish in the same frame.

From the torsion-free Cartan equations,
\be{cartan}
\mathrm{d}e^{\hat a}+\omega^{\hat a}{}_{\hat b}\wedge e^{\hat b}=0,
\ee
we evaluate the Einstein tensor in the orthonormal frame associated with~\eqref{metric:gen}.
Using $R^{\hat a}{}_{\hat b}=\rd\omega^{\hat a}{}_{\hat b}
+\omega^{\hat a}{}_{\hat c}\wedge\omega^{\hat c}{}_{\hat b}$ and the standard conversion to $G_{\hat a\hat b}$, we obtain the mixed components,
\bea
G_{\hat1\hat2} 
&=&\frac{3\sin\theta \sqrt{q \Delta }}{2 \Sigma }\left[\left( \frac{a^2 
   \dot {p}}{\Gamma -a^2 p}\right)'
   -\frac{\dot{\Sigma}\Sigma' }{\Sigma^2}
   + \frac13 \frac{\dot{q}}{q}  \frac{\Sigma'}{\Sigma} 
   	+\frac23 \frac{\dot{\Sigma}' }{\Sigma}\right] , \nn \\
G_{\hat 0\hat 3} &=&-\frac{a \sin \theta 
	\sqrt{q^{3/2}\Delta }}{2\Sigma\left(\Gamma -a^2 p\right)} 
 \Biggl[\Gamma '' 
 	+ \left(\frac{\ddot p}{q^2}- \frac{2\dot p \dot q}{q^3} \right)
	+\Gamma ' \frac{\Sigma '}{\Sigma}
\nn \\
&& \qquad  +\frac{\dot p}{q^2}\frac{\dot \Sigma }{\Sigma}
	-  \frac{\left(\Gamma '\right)^2}{\Gamma-a^2p}
	+\frac{a^2 \dot p^2/q^2}{\Gamma -a^2p}
\Biggr] ,
\label{starting}
\eea
where primes and dots denote derivatives with respect to $r$ and $x\equiv \cos\theta$, respectively.

\section{Characteristic Variables and Integration of $G_{\hat0\hat3}=0$} \label{sec:char-var}
To expose the characteristic structure of the equations~\eqref{key}, it is convenient to introduce variables adapted to the natural prefactors appearing above.
We define
\begin{equation}
\partial_R \equiv \Gamma'(r)\,\partial_r,
\qquad
\partial_z \equiv \frac{\dot p(x)}{q(x)^2}\,\partial_x,
\label{Rz:rx}
\end{equation}
which locally define coordinates $R=R(r)$ and $z=z(x)$.
In these variables, the radial and angular derivatives enter on an equal footing and reveal a characteristic direction generated by
$\partial_R+\partial_z$.
The mixed Einstein equations single out a preferred characteristic foliation.

We also define
\be{GammaR-pz}
\Gamma_R\equiv \partial_R\Gamma,
\qquad
p_z\equiv \partial_z p,
\ee
and the characteristic variables
\be{y-ybar}
y\equiv R-z,
\qquad
\bar y\equiv R+z.
\ee
In terms of $(R,z)$, the condition $G_{\hat0\hat3}=0$ reduces to (for details see App.~\ref{SM:char}.)
\begin{equation}
\frac12\,\partial_R\log \Gamma_R
+\frac12\,\partial_z\log\!\left(\frac{p_z}{q^2}\right)
+\left(\partial_R+\partial_z\right)\log\!\left(\frac{\Sigma}{\Gamma-a^2p}\right)
=0 ,
\label{G03=0}
\end{equation}
where $\Gamma_R\equiv\partial_R\Gamma$ and $p_z\equiv\partial_z p$.
Interpreting \eqref{G03=0} as a first-order differential equation for $\Sigma$, we obtain
\begin{equation}
\Sigma(R,z)
=
\frac{\Sigma_0\,(\Gamma-a^2p)\,\varSigma(R-z)}
{\sqrt{|\Gamma_R(p_z/q^2)|}},
\label{varSigma}
\end{equation}
with $\Sigma_0$ a constant and $\varSigma$ an arbitrary function of the characteristic variable $y=R-z$.

\section{Riccati Reduction of $G_{\hat1\hat2}=0$} \label{sec:Riccati}

Substituting Eq.~\eqref{varSigma} into the second mixed Einstein equation gives a nontrivial closure condition.
Introduce
\be{Hdef}
\mathfrak H(y)\equiv \frac{d}{dy}\log\varSigma(y).
\ee
Then the equation $G_{\hat1\hat2}=0$ reduces to a Riccati equation,
\be{RicH}
\frac{d\mathfrak H}{dy}
-\frac12\,\mathfrak H^2
-\frac12\,F(R,z)\,\mathfrak H
-\frac14\,G(R,z)=0,
\ee
where
\bea
F(R,z) &=&
\frac{\Gamma_R+a^2 p_z}{\Gamma-a^2 p}
+\frac12\left(
-\frac{\Gamma_{RR}}{\Gamma_R}
+\frac{p_{zz}}{p_z}
\right),
\label{Fdef}\\[4pt]
G(R,z) &=&
\frac{p_{zz}}{p_z}\frac{\Gamma_R}{\Gamma-a^2 p}
 - \frac{a^2p_z}{\Gamma-a^2 p}\frac{\Gamma_{RR}}{\Gamma_R}
-\frac{p_{zz}}{2p_z}\frac{\Gamma_{RR}}{\Gamma_R} .
\label{Gdef}
\eea
Since $\mathfrak H$ depends only on $y=R-z$, consistency requires both $F$ and $G$ to be functions of $y$ alone.
Using the standard transformation $\mathfrak H=-2u_y/u$, the Riccati equation becomes the second-order linear equation
\be{Riccati}
\frac{d^2 u}{dy^2}
-\frac12 F\,\frac{du}{dy}
+\frac18 G\,u=0,
\qquad
\varSigma(y)=\frac{\varSigma_0}{u(y)^2}.
\ee
The projective rigidity of the geometry is therefore encoded in the closure conditions that make the coefficients of this equation genuinely characteristic.
The essential point is therefore not merely the Riccati form itself, but the closure requirement that its coefficients become genuine functions of the characteristic variable $y$.

\section{Schwarzian Constraint and Projective Alignment}\label{sec:schwarzian}
Since \(u\) depends only on the characteristic variable \(y=R-z\), consistency requires that both \(F\) and \(G\) be functions of \(y\) alone. This closure condition imposes strong constraints on the metric functions and is the origin of the projective rigidity. In particular, from eq.~\eqref{Fdef} one may write
\begin{equation}
F=\partial_y \log S(R,z),
\qquad
S(R,z)=\frac{(\Gamma-a^2p)^2}{\Gamma_R\,p_z}.
\end{equation}
Therefore the requirement \(F=F(y)\) implies
\begin{equation}
\partial_{\bar y}\partial_y \log S(R,z)=0,
\end{equation}
or equivalently,
\begin{equation}
(\partial_R^2-\partial_z^2)\log S(R,z)=0.
\end{equation}
As shown in App.~\ref{SM:schwarzian}, this closure condition can be reorganized into a compatibility equation for a mixed quantity \(M(R,z)\), from which the equality of Schwarzian derivatives follows.

The resulting consistency condition is
\begin{equation}
\{\Gamma,R\}=\{p,z\}=C,
\label{SchD}
\end{equation}
where
\begin{equation}
\{A,B\}\equiv \frac{A_{BBB}}{A_B}-\frac32\left(\frac{A_{BB}}{A_B}\right)^2
\label{SchAB}
\end{equation}
is the Schwarzian derivative and \(C\) is a constant, since the left-hand side depends only on \(R\) while the right-hand side depends only on \(z\). Equation~\eqref{SchD} is invariant under independent M\"obius transformations of \(\Gamma\) and \(p\), reflecting the projective nature of the closure condition. It therefore fixes the radial and angular sectors up to projective transformations and provides the projective backbone of the resulting separable structure.

Equation~\eqref{SchD} fixes the functions $\Gamma(R)$ and $p(z)$ up to projective transformations.
Accordingly, the general solution is classified into three distinct branches, depending on the sign of the Schwarzian constant $C$,
\begin{equation}
\Gamma(R)=
\begin{cases}
\dfrac{\alpha+\beta R}{\gamma+\delta R},
& C=0,\\[8pt]
\dfrac{\alpha e^{\mu R}+\beta e^{-\mu R}}
{\gamma e^{\mu R}+\delta e^{-\mu R}},
& C<0,\\[8pt]
\dfrac{\alpha\cos(\mu R)+\beta\sin(\mu R)}
{\gamma\cos(\mu R)+\delta\sin(\mu R)},
& C>0,
\end{cases}
\label{GammaR}
\end{equation}
with $\alpha\delta-\beta\gamma\neq0$ and an analogous expression for $p(z)$.
Here $C=\mp 2\mu^2$ for the exponential/trigonometric branch.
For the Kerr family, $\Gamma(r)=r^2+a^2$ and $\partial_R=\Gamma'(r)\partial_r=2r\,\partial_r$, so that
\begin{equation}
\{\Gamma,R\}=-8 .
\end{equation}
Thus Kerr lies on a fixed constant-Schwarzian orbit within the $C<0$ branch in the projective coordinate $R$, providing a concrete realization of the projective class identified above.

The closure condition that the function $F(R,z)$ depend only on the
characteristic variable $y=R-z$ imposes a stronger restriction than
\eqref{SchD}.
Since $F=\partial_y \log S$, consistency requires the alignment of the radial and angular sectors within the same projective parametrization.
As shown in App.~\ref{App:alignment} this is possible if and only if 
\begin{equation}
p(z)=\frac{1}{a^2}\,\Gamma\!\left(\pm(z-z_0)\right),
\label{pGamma}
\end{equation}
where $z_0$ is a constant.
The relation~\eqref{pGamma} defines what we call a \emph{projective alignment} between the radial and angular sectors.
The two choices correspond to a direct alignment and an anti-alignment along the characteristic directions.
Thus the relative projective freedom between $\Gamma(R)$ and $p(z)$ is fully fixed, leaving the Schwarzian constant $C$ as the sole remaining parameter.

With \eqref{pGamma} imposed, the remaining closure condition associated with $G(R,z)$ introduces no further constraints: $G$ reduces to a constant whose value is fixed by the branch and the $\pm$ signature in Eq.~\eqref{pGamma}. 
The Riccati equation \eqref{Riccati} can then be solved explicitly for
$\varSigma(y)$.
While the detailed functional forms are not essential for our discussion, their qualitative behavior plays an important role.

The appearance of the Schwarzian condition is not an artifact of the particular parametrization~\eqref{metric:gen}. 
As shown in App.~\ref{App:robustness}, the same consistency condition persists for a broader stationary and axisymmetric ansatz in which independent radial and angular warp factors are redistributed between the base and fiber sectors. 
This indicates that the projective-Schwarzian rigidity is a genuine consequence of the mixed Einstein equations, rather than a coordinate peculiarity of the chosen form.
Equation~\eqref{SchD} fixes the functional form of $\Gamma(R)$ and $p(z)$ up to M\"obius transformations~\cite{Polyakov81,Lehto87} and provides the projective backbone of the resulting separable structure.

For the M\"obius ($C=0$) and exponential ($C<0$) branches, $\varSigma(y)$ admits elementary solutions that are globally regular for appropriate choices of integration constants.
By contrast, as shown in App.~\ref{SM:trig}, in the trigonometric branch ($C>0$), the solution for $\varSigma(y)$ is expressed in terms of Legendre-type functions of $\cos[\mu(y+z_0)]$ with non-integer degree.
Such functions are generically singular or multi-valued as
$\cos[\mu(y+z_0)]\to -1$, which corresponds to the endpoint of the angular domain (e.g., the rotation axis where regularity is required).
As a result, the trigonometric branch fails to admit globally regular and single-valued geometries compatible with standard periodicity and axis regularity conditions.
We note that even in the exponential branch ($C<0$), some choices of integration constants (and of the sign in \eqref{pGamma}) lead to zeros of $u(y)$ and hence to poles in $\varSigma\propto u^{-2}$ on real $y$.
Requiring regularity of $\varSigma$ on the physical domain therefore restricts the allowed constants (and, in practice, favors the $F=0$ sub-branch), whereas the $C>0$ branch is generically obstructed already by axis periodicity and single-valuedness.
As a result, the trigonometric branch is generically excluded under standard axis regularity and periodicity conditions, whereas the $C\le0$ branches yield well-behaved geometries.

At this stage, the function $\Sigma(R,z)$ is completely fixed up to a finite number of constants.
The remaining freedom resides in the functions $\Delta(r)$ and $q(z)$, which enter only the diagonal components of the Einstein equations and will be fixed by the remaining field equations.

\section{Summary and Discussions}\label{sec:summary}
\vspace{.3cm}
In this work, we have shown that a constant-Schwarzian projective structure arises from the compatibility conditions implied by the mixed Einstein equations in stationary axisymmetric geometries and underlies the local origin of Kerr-type hidden symmetry and separability.

\paragraph{Physical meaning: local equilibrium and rigidity.}
The conditions~\eqref{key} admit a direct physical interpretation in a locally non-rotating orthonormal frame: they exclude local energy and momentum fluxes and thus define a minimal notion of stationary local equilibrium, irrespective of the diagonal components of the stress tensor. 
Remarkably, imposing only this condition already removes most of the functional arbitrariness of a general stationary axisymmetric geometry and drives the Einstein equations into a rigid characteristic form. 
The remaining diagonal equations then determine only the functions $\Delta(r)$ and $q(z)$, showing that, within the present class, the mixed sector controls the emergence of separability and hidden symmetry.

\paragraph{Geometric interpretation: Schwarzian structure and projective alignment.}
The mixed Einstein equations single out the characteristic variable $y=R-z$, reducing the system to a Riccati equation whose closure yields a differential compatibility condition equivalent to the equality of Schwarzian derivatives,
$\{\Gamma,R\}=\{p,z\}$.
This constant-Schwarzian condition fixes the radial and angular sectors up to projective transformations, while the stronger closure requirement enforces the alignment
$p(z)=a^{-2}\Gamma(\pm(z-z_0))$.
The resulting projective alignment places the geometry in an aligned separable Carter-Plebanski-type class, within which Petrov type D algebraic speciality~\cite{Newman1962,Chandrasekhar83}  and the existence of a (principal conformal) Killing-Yano tensor follow in the standard way.

The appearance of the Schwarzian derivative is especially notable because the same structure governs the universal boundary reparametrization mode in nearly-AdS$_2$ holography~\cite{Maldacena2016,Kitaev2018}. 
Our result does not identify Kerr with a two-dimensional theory; rather, it shows that the invariant content of local separability in rotating Einstein geometries is naturally encoded by the same constant-Schwarzian structure that also organizes the universal near-horizon sector of near-extremal Kerr. 
For the Kerr family, the same projective invariant places the near-extremal throat on a constant-Schwarzian orbit, isolating a JT-like sector without implying a full dimensional reduction.

\paragraph{Relation to Kerr uniqueness.}
Our analysis does not constitute a new uniqueness theorem. 
Rather, it identifies a local structural precursor to Kerr uniqueness: before imposing asymptotic flatness, horizon regularity, analyticity, or separability, the Einstein equations already enforce, within the present stationary-axisymmetric class, the kinematical core, the kinematical core of Kerr geometry through projective alignment and constant Schwarzian rigidity. 
Global conditions then act only to select the physically admissible realization within this rigid local class, excluding in particular the trigonometric branch by regularity, periodicity, and single-valuedness requirements.

\paragraph{Near-extremal throat and ringdown relevance.}
This geometric structure also admits a natural near-extremal interpretation.
For the Kerr family, the projective invariant takes the fixed value $\{\Gamma,R\}=-8$, placing the near-extremal throat on a constant-Schwarzian orbit and isolating a JT-like universal sector of the near-horizon dynamics.
The same invariant is also reflected in the Liouville normal form of the radial Teukolsky equation, suggesting a geometric link between the local projective rigidity identified here and the universal low-frequency response of rapidly rotating black holes.
A more detailed analysis of this connection lies beyond the scope of the present work.

A final remark concerns the two realizations of the projective alignment condition,
$p(z)=a^{-2}\Gamma(\pm(z-z_0))$.
Locally, the aligned and anti-aligned branches are equivalent: they obey the same Schwarzian constraint and share the same separable and Petrov type~D structure. 
Their distinction is global. 
In the exponential branch ($C<0$), only the anti-aligned realization 
is the branch naturally compatible with the complex shift leading to the standard globally regular Kerr parametrization.
From this perspective, Kerr is the globally admissible realization of an underlying projectively rigid local class.

Finally, we have verified that the emergence of the Schwarzian consistency condition is robust under broader stationary and axisymmetric parametrizations, including redistributions of independent radial and angular warp factors between the base and fiber sectors. 
This supports the interpretation of the projective-Schwarzian constraint as a genuine consequence of the mixed Einstein equations, rather than a peculiarity of the specific ansatz~\eqref{metric:gen}.

While our analysis is four-dimensional, the projective nature of the constraint suggests possible extensions, which we leave for future work.

\section*{Acknowledgments}
This work was supported by the National Research Foundation of Korea(NRF) grant with grant number RS-2023-00208047 (H.K.).

\appendix
\section{Characteristic variables and integration of $G_{\hat0\hat3}=0$}
\label{SM:char}

\subsection{Definition of $(R,z)$ operators and characteristic coordinates}
Introduce the differential operators~\eqref{Rz:rx},
which locally define new variables $R=R(r)$ and $z=z(x)$ (up to additive constants). We also use Eqs.~\eqref{GammaR-pz} and \eqref{y-ybar} so that
\bea \label{SM:Rz-to-yybar}
\partial_R&=&\partial_y+\partial_{\bar y},
\quad
\partial_z=-\partial_y+\partial_{\bar y}, \nn \\
\partial_R+\partial_z&=&2\partial_{\bar y},
\quad
\partial_R-\partial_z=2\partial_y.
\eea

\subsection{Integration of $G_{\hat0\hat3}=0$}

Using Eq.~\eqref{Rz:rx}, one has the identities
\bea \label{R:r}
\Gamma_R &\equiv& \partial_R\Gamma = \Gamma'(r)\,\partial_r\Gamma = \bigl(\Gamma'(r)\bigr)^2,
\nn\\
p_z &\equiv& \partial_z p = \frac{\dot p}{q^2}\,\dot p = \frac{\dot p^{\,2}}{q^2}, \quad
\Sigma_z \equiv \partial_z\Sigma = \frac{\dot p}{q^2}\,\dot\Sigma.
\eea
Moreover,
\bea
\partial_R\log \Gamma_R
&=& \Gamma'(r)\,\partial_r\log\bigl(\Gamma'(r)\bigr)^2
=2\,\Gamma''(r), \nn \\
\partial_z\log\!\left(\frac{p_z}{q^2}\right)
&=& \frac{\dot p}{q^2}\,\partial_x\log\!\left(\frac{\dot p^{\,2}}{q^4}\right)
=2\left(\frac{\ddot p}{q^2}-\frac{2\dot p\,\dot q}{q^3}\right).
\eea

With these relations, the condition \(G_{\hat0\hat3}=0\) from  Eq.~\eqref{starting} can be rewritten as
\bea
&&\frac12\,\partial_R\log \Gamma_R
	+\frac12\,\partial_z\log\!\left(\frac{p_z}{q^2}\right)
	+\Gamma'\frac{\Sigma'}{\Sigma}
	+\frac{\dot p}{q^2}\frac{\dot \Sigma}{\Sigma}
 \nn \\
&&-\frac{(\Gamma')^2}{\Gamma-a^2p}
+\frac{a^2\dot p^{\,2}/q^2}{\Gamma-a^2p}=0.
\eea
Using \(\Gamma'\Sigma'/\Sigma=\partial_R\log\Sigma\) and \((\dot p/q^2)\Sigma_x/\Sigma=\partial_z\log\Sigma\), together with
\[ -\frac{(\Gamma')^2}{\Gamma-a^2p}
	+\frac{a^2\dot p^{\,2}/q^2}{\Gamma-a^2p}
	=-(\partial_R+\partial_z)\log(\Gamma-a^2p),
\]
we obtain the compact form
\begin{equation}
\boxed{
\frac12\,\partial_R\log \Gamma_R
	+\frac12\,\partial_z\log\!\left(\frac{p_z}{q^2}\right)
	+\left(\partial_R+\partial_z\right)\log\!\left(\frac{\Sigma}{\Gamma-a^2p}\right)=0
}.
\end{equation}

We isolate the characteristic derivative:
\be{SM:G03PDE-iso}
\left(\partial_R+\partial_z\right)\log\!\left(\frac{\Sigma}{\Gamma-a^2p}\right)
= -\frac12\,\partial_R\log \Gamma_R
	-\frac12\,\partial_z\log\!\left(\frac{p_z}{q^2}\right).
\ee
Now note that $\Gamma_R=\Gamma_R(R)$ depends only on $R$, while $(p_z/q^2)=(p_z/q^2)(z)$ depends only on $z$. Using $\partial_R+\partial_z$,
\bea
\left(\partial_R+\partial_z\right)\log\Gamma_R
&=&\partial_R\log\Gamma_R,
\nn \\
\left(\partial_R+\partial_z\right)\log\!\left(\frac{p_z}{q^2}\right)
&=&\partial_z\log\!\left(\frac{p_z}{q^2}\right),
\label{SM:char-identities}
\eea
so Eq.~\eqref{SM:G03PDE-iso} can be rewritten as
\bea
\left(\partial_R+\partial_z\right)\left[
	\log\!\left(\frac{\Sigma}{\Gamma-a^2p}\right)
	+\frac12\log\!\left(\frac{\Gamma_Rp_z}{q^2}\right)
\right]=0.
\label{SM:G03-total-deriv}
\eea
Equivalently, in $(y,\bar y)$ variables,
\be{SM:G03-dybar}
2\partial_{\bar y}\left[
	\log\!\left(\frac{\Sigma}{\Gamma-a^2p}\right)
	+\frac12\log\!\left(\frac{\Gamma_Rp_z}{q^2}\right)
\right]=0,
\ee
which implies that the bracket is an arbitrary function of $y$ alone:
\be{SM:G03-integrated-log}
\log\!\left(\frac{\Sigma}{\Gamma-a^2p}\right)
	+\frac12\log\Gamma_R
	+\frac12\log\!\left(\frac{p_z}{q^2}\right)
=\log\!\left(\Sigma_0\,\varSigma(y)\right).
\ee
Exponentiating yields
\be{SM:SigmaSol}
\Sigma(R,z)
=\frac{\Sigma_0\,(\Gamma-a^2p)\,\varSigma(y)}
	{\sqrt{\Gamma_R\,\left(p_z/q^2\right)}} ,
\qquad y=R-z,
\ee
where $\Sigma_0$ is a constant and $\varSigma$ is an arbitrary real function of the characteristic variable $y$.
In the main text we write the denominator as $\sqrt{\Gamma_R(p_z/q^2)|}$ to allow for sign conventions in $R$ and $z$; this does not affect the subsequent analysis.

\section{Derivation of the Riccati equation from $G_{\hat1\hat2}=0$}
\label{SM:Riccati}

\subsection{Reduced equation for $\varSigma(y)$}

By using the $R$ and $z$ coordinates in Eq.~\eqref{Rz:rx}, the $G_{\hat 1\hat 2}=0$ equation in Eq.~\eqref{starting} can be written as
\be{R12=0}
\left( \frac{a^2 
   p_z}{\Gamma -a^2 p}\right)_R
   =\frac{\Sigma_z\Sigma_R }{\Sigma^2}
   - \frac13 \frac{q_z}{q}  \frac{\Sigma_R}{\Sigma} 
   	-\frac23 \frac{\Sigma_{Rz} }{\Sigma},
\ee
where $()_R \equiv \partial_R ()$. 

Starting from Eq.~\eqref{SM:SigmaSol}, it is convenient to introduce
\begin{equation}
\mathfrak H(y)\;\equiv\;\frac{d}{dy}\log\varSigma(y).
\end{equation}
This definition linearizes the dependence on $\varSigma$ and casts the reduced equation into a standard Riccati form.

Writing
\begin{equation}
\log\frac{\Sigma}{\Sigma_0}
=\log(\Gamma-a^2p)
	-\frac12\log\Gamma_R
	-\frac12\log\!\left(\frac{p_z}{q^2}\right)
	+\log\varSigma(y),
\end{equation}
we obtain
\begin{align}
\partial_R\log\Sigma&=
	\frac{\Gamma_R}{\Gamma-a^2p}
	-\frac12\,\partial_R\log\Gamma_R
	+\mathfrak H(y),\\[4pt]
\partial_z\log\Sigma &=
	-\frac{a^2p_z}{\Gamma-a^2p}
	-\frac12\,\partial_z\log\!\left(\frac{p_z}{q^2}\right)
	-\mathfrak H(y).
\end{align}
Similarly, using \(\Sigma_R=\Sigma\,\partial_R\log\Sigma\), \(\Sigma_z=\Sigma\,\partial_z\log\Sigma\), and
\(\Sigma_{Rz}=\Sigma\bigl[(\partial_R\log\Sigma)(\partial_z\log\Sigma)+\partial_R\partial_z\log\Sigma\bigr]\),
the mixed equation \(G_{\hat1\hat2}=0\) in Eq.~\eqref{starting},
reduces to an equation involving \(\mathfrak H(y)\), together with coefficient functions built from \(\Gamma(R)\) and \(p(z)\).
Explicitly putting these results to Eq.~\eqref{R12=0}, we have a Riccati type equation, 
\begin{equation}
\frac{d\mathfrak H}{dy}
-\frac12\,\mathfrak H^2
-\frac12\,F(R, z)\,\mathfrak H
-\frac14\,G(R, z)=0, 
\end{equation}
where 
\begin{align}\label{F}
F(R,z) &\equiv
\frac{\Gamma_R+a^2 p_z}{\Gamma-a^2 p}
+\frac12\left(
	-\frac{\Gamma_{RR}}{\Gamma_R}
	+\frac{p_{zz}}{p_z}
\right),
\\[4pt]
G(R,z) &\equiv
\frac{p_{zz}}{p_z}\frac{\Gamma_R}{\Gamma-a^2 p}
 - \frac{a^2p_z}{\Gamma-a^2 p}\frac{\Gamma_{RR}}{\Gamma_R}
-\frac{p_{zz}}{2p_z}\frac{\Gamma_{RR}}{\Gamma_R} , \nn 
\end{align}
where the functions $F$ and $G$ must be functions of \(y\) only for the Riccati equation to be consistent.
To solve the differential equation, we use the standard  transformation $\mathfrak{H} =- 2(du/dy)/u$ to get 
\be{App Riccati}
 \frac{d^2 u}{dy^2}  - \frac12 F \frac{du}{dy}+\frac18 G u=0, \qquad
 \varSigma(y) = \frac{\varSigma_0}{u(y)^2}.
\ee

\section{Derivation of the Schwarzian constraint}\label{SM:schwarzian}
In this appendix we derive the Schwarzian consistency condition from the closure requirement that the coefficient \(F(R,z)\) in the Riccati equation depend only on the characteristic variable \(y=R-z\).

\subsection{Factorization condition}
From eq.~\eqref{Fdef}, one verifies that
\begin{equation}
F(R,z)=\partial_y \log S(R,z),
\qquad
S(R,z)\equiv \frac{\bigl(\Gamma(R)-a^2p(z)\bigr)^2}{\Gamma_R\,p_z},
\label{AppE:FlogS}
\end{equation}
where, consistently with eq.~\eqref{SM:Rz-to-yybar},
\begin{equation}
\partial_y=\frac12(\partial_R-\partial_z),
\qquad
\partial_{\bar y}=\frac12(\partial_R+\partial_z).
\end{equation}
Since \(F\) must be a function of \(y\) only, its \(\bar y\)-derivative must vanish:
\begin{equation}
\partial_{\bar y}F
=
\partial_{\bar y}\partial_y \log S(R,z)=0.
\end{equation}
Using the definitions above, this is equivalent to
\begin{equation}
\left(\partial_R^2-\partial_z^2\right)\log S(R,z)=0.
\label{AppE:keyeq}
\end{equation}

\subsection{Direct derivation of the Schwarzian condition}
Now define
\begin{equation}
Q(R,z)\equiv \Gamma(R)-a^2p(z),
\end{equation}
so that
\begin{equation}
\log S = 2\log Q-\log \Gamma_R-\log p_z .
\end{equation}
The first and second derivatives are
\begin{align}
\partial_R \log S
&=
\frac{2\Gamma_R}{Q}
-\frac{\Gamma_{RR}}{\Gamma_R},
\qquad
\partial_R^2 \log S
=
\frac{2\Gamma_{RR}}{Q}
-\frac{2\Gamma_R^2}{Q^2}
-\left(
\frac{\Gamma_{RRR}}{\Gamma_R}
-\frac{\Gamma_{RR}^2}{\Gamma_R^2}
\right),
\nn \\
\partial_z \log S
&=
-\frac{2a^2p_z}{Q}
-\frac{p_{zz}}{p_z},
\qquad
\partial_z^2 \log S
=
-\frac{2a^2p_{zz}}{Q}
-\frac{2a^4p_z^2}{Q^2}
-\left(
\frac{p_{zzz}}{p_z}
-\frac{p_{zz}^2}{p_z^2}
\right).
\end{align}
Substituting these expressions into eq.~\eqref{AppE:keyeq}, we obtain
\begin{equation}
M(R,z)
\equiv
\frac{\Gamma_{RR}+a^2p_{zz}}{Q}
-\frac{\Gamma_R^2-a^4p_z^2}{Q^2}
=
\frac12
\left[
\left(\frac{\Gamma_{RR}}{\Gamma_R}\right)_R
-
\left(\frac{p_{zz}}{p_z}\right)_z
\right].
\label{AppE:Mdef}
\end{equation}

The right-hand side of eq.~\eqref{AppE:Mdef} is a sum of a function of \(R\) and a function of \(z\). Therefore, compatibility requires
\begin{equation}
\partial_R\partial_z M=0.
\label{AppE:compatM}
\end{equation}
On the other hand, from the definition of \(M\),
\begin{equation}
M
= \partial_R\!\left(\frac{\Gamma_R}{Q}\right)
-\partial_z\!\left(\frac{-a^2p_z}{Q}\right).
\end{equation}
A direct computation then gives
\begin{align}
\partial_R\partial_z M
&=
\frac{a^2\Gamma_R p_z}{Q^4}
\left[
Q^2\left(
\frac{\Gamma_{RRR}}{\Gamma_R}
-\frac{p_{zzz}}{p_z}
\right)
-6(\Gamma_{RR}+a^2p_{zz})Q
+6(\Gamma_R^2-a^4p_z^2)
\right]
\nn \\
&=
\frac{a^2\Gamma_R p_z}{Q^2}
\left[
\left(
\frac{\Gamma_{RRR}}{\Gamma_R}
-\frac{p_{zzz}}{p_z}
\right)
-6M
\right].
\label{AppE:dRdzM}
\end{align}
Combining eqs.~\eqref{AppE:compatM} and \eqref{AppE:dRdzM}, we find
\begin{equation}
\frac{\Gamma_{RRR}}{\Gamma_R}
-\frac{p_{zzz}}{p_z}
=6M.
\label{AppE:Xeq6M}
\end{equation}
Substituting the definition of \(M\) from eq.~\eqref{AppE:Mdef}, this becomes
\begin{equation}
\frac{\Gamma_{RRR}}{\Gamma_R}
-\frac{p_{zzz}}{p_z}=
3\left[
\left(\frac{\Gamma_{RR}}{\Gamma_R}\right)_R
-
\left(\frac{p_{zz}}{p_z}\right)_z
\right]
=
3\left[
\frac{\Gamma_{RRR}}{\Gamma_R}
-\frac{\Gamma_{RR}^2}{\Gamma_R^2}
-\frac{p_{zzz}}{p_z}
+\frac{p_{zz}^2}{p_z^2}
\right].
\end{equation}
Rearranging terms yields
\begin{equation}
\frac{\Gamma_{RRR}}{\Gamma_R}
-\frac32\left(\frac{\Gamma_{RR}}{\Gamma_R}\right)^2
=
\frac{p_{zzz}}{p_z}
-\frac32\left(\frac{p_{zz}}{p_z}\right)^2 .
\end{equation}
Therefore the Schwarzian derivatives are equal:
\begin{equation}
\{\Gamma,R\}=\{p,z\},
\qquad
\{A,B\}\equiv
\frac{A_{BBB}}{A_B}
-\frac32\left(\frac{A_{BB}}{A_B}\right)^2.
\label{AppE:Schwarzian}
\end{equation}
Since the left-hand side depends only on \(R\), while the right-hand side depends only on \(z\), both must be equal to a constant,
\begin{equation}
\{\Gamma,R\}=\{p,z\}=C.
\label{SM:SchwarzEq}
\end{equation}

It is convenient to introduce
\begin{equation}
U(R)=\frac{1}{\sqrt{\Gamma_R}},
\qquad
V(z)=\frac{1}{\sqrt{p_z}}.
\end{equation}
Then the Schwarzian derivatives take the form
\begin{equation}
\{\Gamma,R\}=-2\frac{U_{RR}}{U},
\qquad
\{p,z\}=-2\frac{V_{zz}}{V},
\end{equation}
so that the constant-Schwarzian condition reduces to
\begin{equation}
U_{RR}=\lambda U,
\qquad
V_{zz}=\lambda V,
\qquad
\lambda=-\frac{C}{2}.
\end{equation}
This yields the three projectively inequivalent branches discussed in the main text: the M\"obius branch for \(C=0\), the exponential branch for \(C<0\), and the trigonometric branch for \(C>0\).

\paragraph{Remark 1.}
Equation~\eqref{SM:SchwarzEq} is invariant under independent M\"obius transformations of $\Gamma$ and $p$, reflecting the projective nature of the closure condition. 
This invariance underlies the three-branch classification in the main text.

\paragraph{Remark 2.}
Since $u$ depends only on $y$, consistency requires that both $F$ and $G$ be functions of $y$ alone, which is the integrability condition that ensures the existence of a generic solution $u(y)$ (equivalently, an arbitrary $\varSigma(y)$). 
If one demands only a particular nontrivial solution, e.g.\ $u\propto e^{\lambda y}$, this requirement can be relaxed to the pointwise constraint $G(R,z)=4\lambda F(R,z)-8\lambda^2$; such cases therefore represent a special sub-class rather than the generic branch discussed here.

\section{Solutions of constant-Schwarzian equation and three branches}
\label{App:alignment}

\subsection{General solution for $\Gamma(R)$}
Solving $\{\Gamma,R\}=C$ yields three projectively inequivalent families:
\be{SM:GammaBranches}
\Gamma(R)=
\begin{cases}
\dfrac{\alpha+\beta R}{\gamma+\delta R},
& C=0,\\[8pt]
\dfrac{\alpha e^{\mu R}+\beta e^{-\mu R}}
{\gamma e^{\mu R}+\delta e^{-\mu R}},
& C=-2\mu^2<0,\\[8pt]
\dfrac{\alpha\cos(\mu R)+\beta\sin(\mu R)}
{\gamma\cos(\mu R)+\delta\sin(\mu R)},
& C=+2\mu^2>0,
\end{cases}
\ee
with $\alpha\delta-\beta\gamma\neq0$.
An analogous expression holds for $p(z)$.

\subsection{Alignment condition $p(z)=a^{-2}\Gamma(\pm(z-z_0))$}
Imposing the stronger closure requirement that $F$ depend only on $y=R-z$ fixes the relative projective freedom, leading to
\be{SM:pGamma}
p(z)=\frac{1}{a^2}\,\Gamma\!\left(\pm(z-z_0)\right),
\ee
with constant $z_0$.

To show the result, we illustrate the M\"{o}bius ($C=0$) case as an example.
Similar calculations lead the same results for other cases also. 
As a solution of Schwarzian equation, $\Gamma$ and $p$ take the form, 
$$
\Gamma 
= \frac{\alpha +\beta R}{\gamma + \delta R}, \qquad
p =   \frac{\alpha' +\beta' z}{\gamma' + \delta' z},
$$
where $\alpha'$, $\beta'$, $\gamma'$, and $\delta'$ are independent of $\alpha$, $\beta$, $\gamma$, and $\delta$ at the present state. 
Until now, we have shown that when $\frac{(\Gamma(R)-a^2 p(z))^2}{\Gamma_R p_z}$ is separable into the multiplication of two functions to the form, $ \tilde F(y) \tilde G(\bar y)$, the Schwarzschian derivatives of $\Gamma$ and $p$ are constant.  
However, the converse never hold automatically. 

Let us check how the separable condition holds. 
\bea
\Gamma_R &=& \frac{\beta \gamma-\delta\alpha}
			{(\gamma+ \delta R)^2}, \qquad
p_z = \frac{\beta' \gamma'-\delta'\alpha'}
			{(\gamma'+ \delta' z)^2},
\nn \\
\Gamma- a^2 p&=& \frac{(\alpha+ \beta R) (\gamma' + \delta' z) 
				+ a^2(\alpha'+ \beta' z)(\gamma+ \delta R)}
			{(\gamma + \delta R)(\gamma' + \delta' z) }
\eea
Putting this result and using the definition of $y$ and $\bar y$ in Eq.~\eqref{y-ybar}, we get 
\bea
&&\frac{(\Gamma(R)-a^2 p(z))^2}{\Gamma_R p_z} 
= \frac1
	{(\beta\gamma-\delta\alpha)(\beta'\gamma'-\delta'\alpha')} 
\times
\nn \\
&&~\left[\frac{\beta\delta' - a^2\beta'\delta}{4}( \bar y^2 -y^2) 
 	+\frac{\alpha\delta'+\beta\gamma'
			-a^2(\alpha'\delta+ \beta'\gamma) }{2} \bar y
	\right. \nn \\
&& \left.~+\frac{-\alpha \delta'+\beta\gamma'
				-a^2(\alpha'\delta -\beta'\gamma) }{2}y 
				+ \alpha \gamma' -a^2\alpha'\gamma\right]^2.
\eea
From this, we find that the function $\frac{(\Gamma(R)-a^2 p(z))^2}{\Gamma_R p_z}$ cannot be factorable into non-degenerate functions but one of $\tilde F(y)$ or $\tilde G(\bar y)$ must be a constant. 
Therefore, when the function is only function of $y$, we have
$$
\beta\delta' = a^2\beta'\delta , \qquad
\alpha \delta'+\beta\gamma'
			=a^2(\alpha'\delta+ \beta'\gamma).
$$
When the function is only function of $\bar y$, we have
$$
\beta\delta' = a^2\beta'\delta , \qquad
-\alpha \delta'+\beta\gamma'
				=a^2(\alpha'\delta -\beta'\gamma) .
$$

For the first, case, given $\Gamma 
= \frac{\alpha +\beta R}{\gamma + \delta R}$.  
Because the scaling $\alpha'$, $\beta'$, $\gamma'$, and $\delta'$ by the same factor does not affect on the function $p$ itself, 
we choose $\beta' = \beta/a^2$ and $\delta' = \delta$.
The function $p(z)$ becomes
$$
p =   \frac{\alpha' +\beta/a^2\, z}{\gamma' + \delta z},
$$
Putting this result, the second condition becomes
$$
\gamma'=(a^2\alpha'-\alpha)\frac{\delta}{\beta}+ \gamma 
$$
Therefore, $p(z)$ becomes, with only one new parameter $\alpha'$, 
$$
p(z) 
	= \frac{1}{a^2}  \frac{\alpha  
		+ \beta ( z+ \frac{\alpha'-\alpha}{\beta})}
		{\gamma + \delta ( z+ \frac{\alpha'- \alpha}{\beta} ) }
	= \frac{\Gamma(z-z_0)}{a^2} ,
$$
where in the second equality we set $a^2\alpha' \to \alpha'$ and then $z_0 = (\alpha-\alpha')/\beta$.

For the second case, the first constraint is the same.
The second condition gives 
$$
\gamma'
=\frac{\alpha \delta-\beta\gamma}{\beta}+ a^2\alpha'\frac{\delta}{\beta} 
$$
Therefore, $p(z)$ becomes, with only one new parameter $\alpha'$, 
$$
p(z) 
 = \frac{1}{a^2}  \frac{\alpha
 		+\beta \left(- z - \frac{\alpha+\alpha'}{\beta}\right) }
 { \gamma + \delta \left( -z - \frac{\alpha+\alpha'}{\beta} \right)}
 = \frac{\Gamma(z_0 -z)}{a^2}
$$
where we set $a^2\alpha'  \to \alpha' $ in the second equality and $z_0 = -\frac{\alpha+\alpha'}{\beta}$.

The same conclusion extends to the exponential and trigonometric branches because the factorization constraints reduce the independent projective parameters in exactly the same way.

Finally, when we put the resulting functional form for $\Gamma$ and $p$ to the function $G(R, z)$ in Eq.~\eqref{F}, we get $G(R,z)$ takes a constant values $0$, $\pm 2\mu^2$,  and $\mp 2\mu^2$, respectively for $C=0$, $C < 0$, and $C> 0$.

\section{Global behavior of the trigonometric branch}
\label{SM:trig}

\subsection{Reduction to Legendre-type equation}
For the $C>0$ branch, with $s\equiv \mu(y+z_0)$, Eq.~\eqref{App Riccati} can be put into the form
\be{SM:trig-u}
\frac{\rd^2 u}{\rd s^2} + \cA(s)\,\frac{\rd u}{\rd s}+\cB(s)\,u=0,
\ee
where $\cA(s)$ and $\cB(s)$ are explicit functions of $\cos s$.
After the standard transformation $t=\cos s$, the equation reduces to a Legendre-type equation
\be{SM:LegendreEq}
(1-t^2)\,u_{tt}-2t\,u_t+\ell(\ell+1)\,u=0,
\ee
with non-integer $\ell$ (e.g. $\ell=-\frac12\pm\frac{\sqrt2}{2}$ in the main text normalization).

\subsection{Endpoint singularity and multi-valuedness}
The general solution is
\be{SM:uLegendreSol}
u(t)=C_1 P_\ell(t)+C_2 Q_\ell(t),
\qquad t=\cos s,
\ee
so that
\be{SM:varSigmaLegendre}
\varSigma(y)=\frac{\varSigma_0}{u(\cos s)^2}.
\ee
As $t\to -1$ (i.e. $\cos s\to -1$), corresponding to an endpoint of the angular domain where axis regularity and single-valuedness are required, $Q_\ell(t)$ has a logarithmic divergence and $P_\ell(t)$ is generically incompatible with global single-valuedness for non-integer $\ell$, where regularity at one endpoint does not generally extend to a globally single-valued $2\pi$-periodic solution on the full angular domain.
Therefore, the $C>0$ branch is generically excluded by standard global regularity/periodicity requirements.
This divergence cannot be removed by a projective redefinition and therefore signals a genuine geometric obstruction rather than a coordinate artifact.

\section{Robustness of the projective-Schwarzian constraint} \label{App:robustness}
\subsection{Alternative stationary-axisymmetric ansatz}
To test whether the Schwarzian consistency condition is tied to the specific parametrization used in the main text, we consider a more general stationary and axisymmetric ansatz in which additional radial and angular warp factors are redistributed between the base and fiber sectors:
\bea
ds^2 &=& -\frac{\Sigma\Delta}{q(\Gamma-a^2p)^2}(dt-ap\,d\phi)^2
+ \frac{\Sigma}{q\Delta}dr^2 
+ \frac{\Sigma}{QB}d\theta^2
+ \frac{\Sigma Q\sin^2\theta}{B(\Gamma-a^2p)^2}(\Gamma d\phi-a dt)^2.
\eea
This form is inequivalent to Eq.~\eqref{metric:gen} by conformal rescaling unless both $Q$ and $B$ are constants.

\subsection{Persistence of the Schwarzian consistency condition}
Evaluating the Einstein tensor in the locally non-rotating orthonormal frame, we find that the mixed components $G_{\hat0\hat3}$ and $G_{\hat1\hat2}$ retain the same characteristic structure as in the main text.
Imposing the no-flux condition again yields a first-order system whose closure requires
\begin{equation}
\{\Gamma,R\}=\{p,z\},
\end{equation}
where the modified projective coordinate is now defined by
\begin{equation}
\partial_R=\frac{\Gamma'}{B^2}\,\partial_r.
\end{equation}
Thus the emergence of the Schwarzian consistency condition is insensitive to the detailed redistribution of radial and angular warp factors, confirming that the projective rigidity identified in the main text is a genuine consequence of the mixed Einstein equations rather than an artifact of the particular ansatz~\eqref{metric:gen}.


\end{document}